# Imaging the Wigner Crystal of Electrons in One Dimension

I. Shapir[†1], A. Hamo[†1,2], S. Pecker[1], C. P. Moca[3,4], Ö. Legeza[5], G. Zarand[3] and S. Ilani[1*]

[1]*Department of Condensed Matter Physics, Weizmann Institute of Science, Rehovot 76100, Israel.*
[2]*Present address: Department of Physics, Harvard University, Cambridge, Massachusetts 02138, USA.*
[3]*Institute of Physics, Budapest University of Technology and Economics, H-1521 Budapest, Hungary.*
[4]*Department of Physics, University of Oradea, 410087, Romania.*
[5]*Wigner Research Centre for Physics, H-1525, Hungary*
[†] These authors contributed equally to this work
[*] Correspondence to: shahal.ilani@weizmann.ac.il

**The quantum crystal of electrons, predicted more than eighty years ago by Eugene Wigner, is still one of the most elusive states of matter. Here, we present experiments that observe the one-dimensional Wigner crystal directly, by imaging its charge density in real-space. To measure this fragile state without perturbing it, we developed a new scanning probe platform that utilizes a pristine carbon nanotube as a scanning charge perturbation to image, with minimal invasiveness, the many-body electronic density within another nanotube. The obtained images, of few electrons confined in one-dimension, match those of strongly interacting crystals, with electrons ordered like pearls on a necklace. Comparison to theoretical modeling demonstrates the dominance of Coulomb interactions over kinetic energy and the weakness of exchange interactions. Our experiments provide direct evidence for this long-sought electronic state, and open the way for studying other fragile interacting states by imaging their many-body density in real-space.**

Interacting electrons pay an energetic price for getting near each other. If all electrons had similar quantum numbers (e.g. similar spins), this price would be largely reduced because the Pauli exclusion principle keeps like particles apart. Wigner's original observation[1] was that since Pauli exclusion does not operate between electrons of different flavor (spin/valley), they must find a different way to separate in real space to reduce this energetic cost. He predicted that when long-ranged Coulomb interactions dominate over kinetic energy and disorder, a new crystalline ground state should emerge in which the electrons are kept apart irrespective of their flavor. This quantum crystal, which exists even in the absence of flavor, was experimentally searched for in the cleanest available electronic systems, primarily on the surface of liquid helium and in low-dimensional semiconductor heterostructures. On liquid helium, electrons were shown to form classical crystals[2], but due to an inherent instability could not reach the quantum regime. In semiconducting two dimensional electronic systems, transport[3,4], microwave[5,6], NMR[7], optical[8,9], tunneling[10], and bilayer correlations[11] measurements provided indications for the existence of a crystal at high magnetic fields. In one dimension, thermal and quantum fluctuations destroy the long range order, and exclude the crystalline state in an infinite system. However, in finite systems quasi-long range order produces crystalline correlations, and this one-dimensional Wigner crystal state was studied extensively theoretically[12–15] and probed experimentally via transport measurements[16,17]. All experiments performed to date, however, probed only macroscopic properties of this state.

The unambiguous fingerprint of a Wigner crystal lies in its real-space structure, which could in principle be observed with a suitable imaging tool. Early scanning probe experiments imaged the electronic wavefunctions in carbon nanotubes[18–20] (NT) deposited on metallic substrates that screened their electronic interactions, and indeed showed nice agreement with a non-interacting picture. Subsequent experiments reduced the screening by using dielectric substrates or suspended NTs (/nanowires) and observed single electron charging physics induced by the scanning tip[21–23]. These measurements highlighted the inherent difficulty of imaging interacting electrons with conventional scanning methods: To resolve individual electrons, a macroscopic, metallic or dielectric tip should approach the electrons closer than their mutual separation, inevitably screening their interactions that are at the heart of the interacting state. Moreover, macroscopic tips generically carry uncontrollable charges that have strong gating effects on an interacting electron system, strongly distorting the state under study. To image an interacting state, a new kind of scanning probe is therefore needed.

In this work, we introduce a conceptually new scanning probe platform, which utilizes a carbon nanotube as a highly-sensitive, yet minimally-invasive scanning probe for imaging the many-body density of strongly interacting electrons. The miniature size of the probe and the control over the number of its excess electrons allows it to act as a minimal potential perturbation, which can be scanned along another nanotube and image its charge distribution. Using this platform we image the density distribution of few confined electrons, and demonstrate that they form a strongly-interacting Wigner crystal. Detailed theoretical calculations reveal that in the observed crystals, the potential energy largely dominates over the kinetic energy, the exchange interactions are small, and the electrons are separated few times their zero-point motion, making it one of the most extreme interacting states to be measured to date in the solid state.

Our platform comprises a custom-made scan probe microscope, operating at cryogenic temperatures (~10mK), in which two oppositely-facing NT-devices can be brought to close proximity (~100nm)[24], and scanned along each other (Fig. 1a). One device hosts the system-NT, which is used as the one-dimensional system under study (Fig. 1a, bottom). The second device contains the probe-NT, which is perpendicular to the system-NT, and can be scanned along it (Fig. 1a, top). The two devices are assembled using our nano-assembly technique[25], which yields pristine NTs suspended above an array of metallic gates. In the system-NT, it is essential to maintain strong interactions and low disorder, both crucial for obtaining a Wigner crystal. This is achieved by suspending the NT far above the metallic gates ($400\ nm$), thus avoiding their disorder and screening. Using ten electrically-independent gates we design a potential that confines the electrons between two barriers, $\sim 1\ \mu m$ apart, localizing them to the central part of a long suspended nanotube ($L = 2.3\mu m$) (Fig. 1b), away from the contacts that produce undesirable distortions (screening, image charges, disorder, and band bending, see Supp. Info. S1). We use highly opaque barriers that prevent hybridization of the confined electron's wavefunction with those of the electrons in the rest of the NT. Since transport in this situation is highly suppressed, we probe the confined electrons using a charge detector located on a separate segment of the same NT (Fig. 1b, purple). The addition of these electrons are detected as a small change in the detector's current, $I_{CD}$, flowing between the two outer contacts of the device (blue arrow, Fig. 1b) and not through the central segment of the system-NT. The probe-NT device has an almost identical structure, differing only by the NT suspension length ($1.6\ \mu m$) and number of gates (3). Since the probe gates and contacts are perpendicular to the system-NT, the potential they induce remains translationally invariant as the probe is scanned along the system, and the only moving perturbation comes from the moving NT itself (Supp. Info. S2).

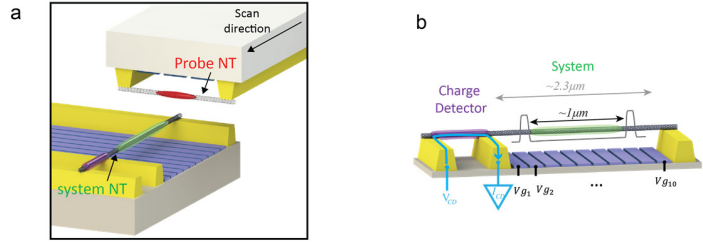

**Figure 1: Experimental platform for imaging strongly-interacting electrons.** a) Scanning probe setup consisting of two carbon nanotube (NT) devices - a 'system-NT' device (bottom) hosts the electrons to be imaged (green ellipse) and a 'probe-NT' device (top) containing the probing electrons (red). In the experiment the probe-NT is scanned along the system-NT (black arrow). b) The system-NT is connected to contacts (yellow) and is suspended above ten gates (blue) used to create a potential well (schematically in gray) that confines few electrons to the middle part of the suspended NT (green), away from the contacts. The charging of these electrons is detected using a charge detector – a separate quantum dot formed on side segment of the same NT (purple). The detector is biased by a voltage $V_{CD}$ applied on an external contact, leading to a current, $I_{CD}$, flowing only between the contacts of the charge detector (blue arrow), such that no current passes through the main part of the system-NT.

To demonstrate the basic principle behind our imaging technique, which we term 'scanning charge', we start with the simplest experiments – imaging the charge distribution of a *single* electron confined to a one-dimensional box (Fig. 2a). The basic idea is that by measuring the energetic response of the system to a scanned perturbation we can directly determine its density distribution. For simplicity, we assume first that the perturbation produced by the probe-NT is highly localized at its position, $x_{probe}$, $V(x) \approx V\delta(x - x_{probe})$. To the lowest order, such a perturbation will shift the system's energy as:

$$E_1(x_{probe}) = \langle\psi_1|V(x - x_{probe})|\psi_1\rangle \propto \rho_1(x_{probe}) \tag{1}$$

where $\rho_1(x) = |\psi_1(x)|^2$ (Fig. 2a, green), is the density distribution of the confined electron wave function, $\psi_1(x)$. Thus, by measuring $E_1$ as a function of $x_{probe}$, the electron's density profile could be directly resolved[26,27]. The energy $E_1$ is measured by referencing it to the Fermi energy in the leads ($E_F \equiv 0$, Fig. 2a bottom). Starting with the level in resonance with $E_F$, a movement of the probe to $x_{probe}$ will shift the energy level to $E_1(x_{probe})$ and generically away from resonance with $E_F$. We then bring the level back to resonance by applying a global gate voltage, $V_g$, formed by a proper linear combination of the ten gates, chosen to produce a rigid shift of the potential without changing its shape (Supp. Info. S1). The applied $V_g$, measured in energy units, is equal to the energy shift $E_1$. Thus, by monitoring the gate voltage required to keep the level in resonance for varying $x_{probe}$, we directly image the charge distribution, $V_g(x_{probe}) \sim \rho_1(x_{probe})$. Note that, in reality, the perturbation produced by our probe is not a delta function, but has a spatial extent determined by the separation between the two nanotubes and the spatial extent of the confined charge in the probe-NT. The measured profile will therefore be the convolution of the corresponding point-spread function and the imaged charge density distribution.

The imaging of the charge distribution of a single electron is shown in Figs. 2b,c. The population of the first confined electron in the system ($q_{system} = 1e$) is identified by its sharp charging peak, observed when we measure the charge detector signal, $dI_{CD}/dV_g$, a function of $V_g$ (Fig. 2b, Supp. Info. S3). To image the charge distribution of this electron, we place one electron also in the probe-NT ($q_{probe} = 1e$), scan it along the system-NT, and monitor the corresponding shift of the charging peak in gate voltage. This measurement (Fig. 2c) reveals that the charging peak shifts smoothly when $x_{probe}$ traverses between the two edges of the confinement well, peaking at its center. This trace yields directly the real-space charge distribution of the first confined electron, convolved with the point spread function of the probe.

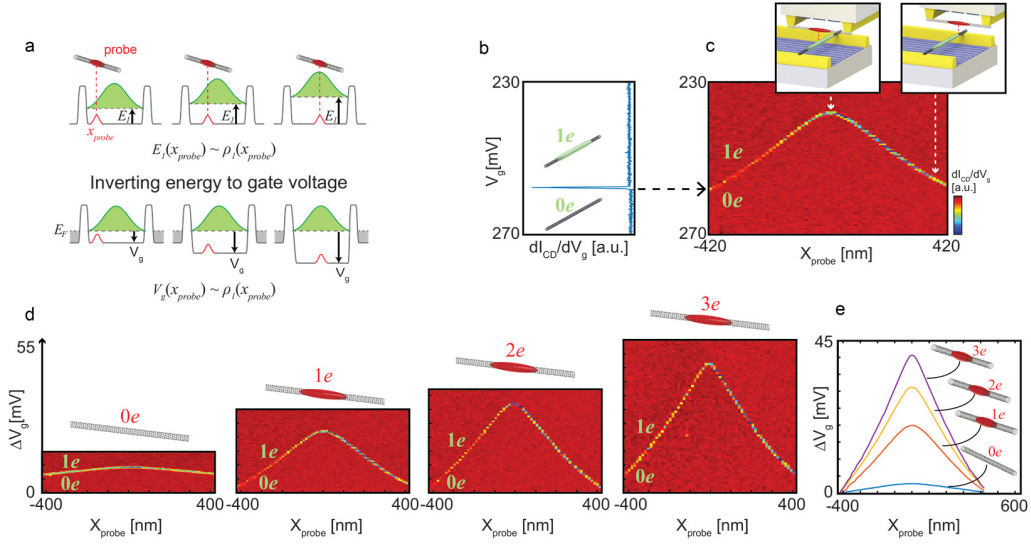

**Figure 2: Real-space imaging of the density profile of a single confined electron.** a) The principle behind our 'scanning charge' imaging technique: To image the density distribution of a single electron confined in a potential 'box' (gray), we place a fixed charge in the probe-NT and scan it across the system-NT. This charge creates a local perturbation at the probe position $x_{probe}$ (red), which shifts the ground state energy of the system electron, $E_1$ (top panels), proportional to the local density at the probe position $E_1(x_{probe}) \sim \rho_1(x_{probe})$ (see text). By measuring the global gate voltage, $V_g$, needed to keep the charging of this single electron in resonance with the Fermi energy of the leads, $E_F$, for varying $x_{probe}$ (bottom panels) we effectively trace the profile of its charge distribution $V_g(x_{probe}) \sim \rho_1(x_{probe})$ b) The derivative of the charge detector current, $dI_{CD}/dV_g$, with respect to $V_g$ measured as a function $V_g$. The sharp charging peak corresponds to the first electron entering the system-NT potential well (throughout the paper the green/red labels mark the number of electrons in the system/probe) c) $dI_{CD}/dV_g$ as a function of $V_g$ and $x_{probe}$. The charging resonance traces a curve that gives the charge density of the electon convolved with the point spread function of the probe. Upper insets: illustration of the system and probe devices for different measurements positions. d) same as in (c) but for different probe charges from $q_{probe} = 0e$ to $q_{probe} = 3e$. e) The traces extracted from panel (d) plotted together. Note that to avoid unnecessary confusion in describing the physics of the electronic Wigner crystal we used throughout the paper the language of electrons, although owing to technical advantages the actual measurements were done with holes in the system and probe (Supp. Info. S3 for details).

An essential test for the technique is to assess how the measured energy shifts scale with the strength of the scanned perturbation, which we can control down to the single electron limit. Figure 2d shows imaging measurements done with $q_{probe} = 0e$ to $3e$, showing that the energy shifts increase monotonically with increasing number of probing electrons. While the overall shape of the imaged charge density remains similar, for larger $q_{probe}$ the peak becomes slightly sharper reflecting an increasing probe invasiveness. By directly measuring the movement of the electron in the system due to the scanned probe (Supp. Info. S4) we see that for measurements done with $q_{probe}$=1 this movement is about an order of magnitude smaller than the zero point motion of this electron, putting them in the non-invasive limit. In principle, our probe could have also had uncontrolled charges, due to localized states in the NT or imperfection in the metals, that can create an even larger scanning perturbation than the single charges that we place intentionally. However, our measurement with $q_{probe} = 0e$ displays an energy shift an order of magnitude smaller than that with $q_{probe} = 1e$, demonstrating that in our experiment the spurious charges are much less significant than one electron.

Having established our imaging technique on a single electron, we now turn to image the interacting states of many electrons. The measurement is similar to the one described above, but is now done around the charging resonance of the $Nth$ electron ($N > 1$, Fig. 3a). This resonance occurs when the states with $N$ and $N-1$ electrons are energetically degenerate, $E_N = E_{N-1}$ (equivalently, $\mu \equiv E_N - E_{N-1} = 0$). The probe perturbation can modify either $E_N$ or $E_{N-1}$, shifting the resonance in gate voltage as:

$$V_g(x_{probe}) = E_N(x_{probe}) - E_{N-1}(x_{probe}) \propto \rho_N(x_{probe}) - \rho_{N-1}(x_{probe}), \quad (2)$$

The trace $V_g(x_{probe})$ now images the density of the $N$ electron state, $\rho_N(x)$, minus that of the $N-1$ electron state, $\rho_{N-1}(x)$. This quantity, which we term the '*differential density*' is intuitive to understand within the single particle picture – it is merely the density added by the last electron to enter the system.

Along the lines of Wigner's argument, when electrons have multiple internal 'flavors' (in our case spin and valley), their differential density profiles should be markedly different in the non-interacting and the strongly-interacting cases. In the absence of interactions, electrons populate the particle-in-a-box states of the potential, with a degeneracy given by the number of flavors. In NTs, this would be a four-fold degeneracy due to the spin and valley degrees of freedom. Thus, the spatial distribution of the density added by each one of the first four electrons (their differential density) would be identical, given by the first particle-in-a-box state. The next four electrons will also look identical, having the two spatial peaks of the second particle-in-a-box state, and so on (Fig. 3b, left). An utterly different picture emerges in the presence of strong interactions: Since Coulomb interactions are flavor independent, all electrons would keep apart independent of their flavor. Thus with any additional electron, one more peak will be added to the differential density profile (Fig. 3b, right). Imaging the differential density of 'flavored' electrons should therefore make a clear distinction between these two regimes.

The imaging of the differential density of the many electron states is shown in Fig. 3c, where the six panels correspond to the first six electrons added to the system-NT. To keep the perturbation minimal, all these scans are performed with one electron in the probe-NT. Different than Fig. 2c, here we plot the charge detector current, $I_{CD}$, rather than its derivative, which shows a step rather than a peak when an electron is added to the system-NT. A clear trend can be observed in the imaged differential density profiles – with every added electron, one more peak appears in the differential density. These profiles are clearly different than those predicted by single particle physics, but nicely match those of a strongly interacting crystal. With increasing number of electrons we see that the electron spacing is reduced, but also that their overall spread increases, signifying that they are confined to a 'box' with soft walls. The slight deviation from perfect periodicity stems from non-ideality of the potential, explained in Supp. Info. S5. Overall, these images are the first direct images of small electronic Wigner crystals.

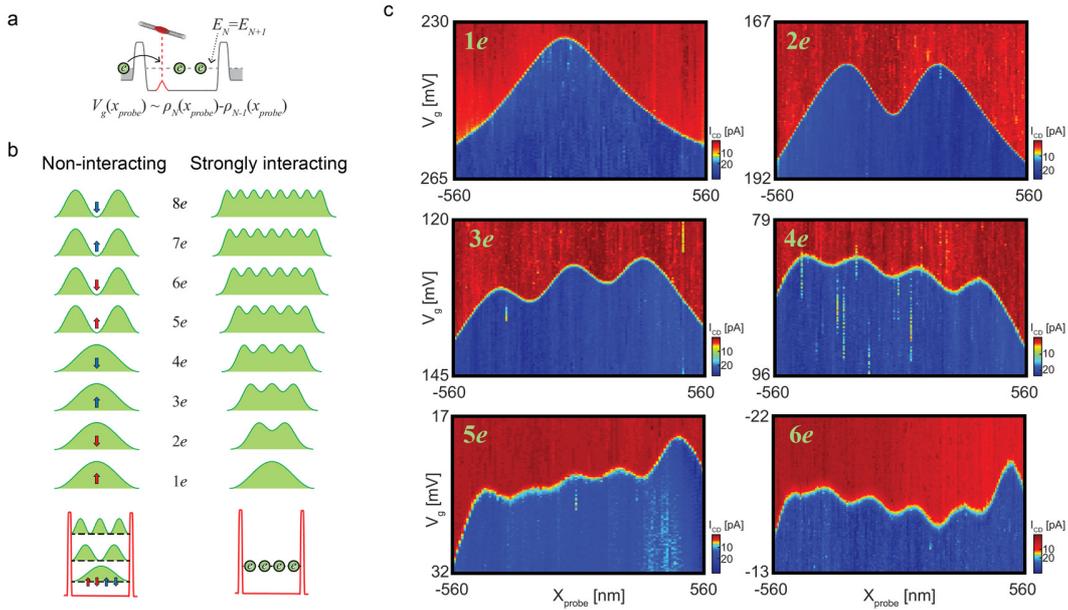

**Figure 3: Imaging the differential density of many-electron states.** a) Similar to the measurements of a single electron (Fig. 2a), here we also probe a charging transition, but now from $N-1$ to $N$ electrons, with the resonance occurring for $E_N = E_{N-1}$. Consequently the gate voltage shifts will now image the differential density $V_g(x_{probe}) \sim \rho_N(x_{probe}) - \rho_{N-1}(x_{probe})$ (see text). b) Illustration of the expected differential density of non-interacting vs. strongly-interacting electrons in a carbon NT. Non-interacting electrons occupy the particle-in-a-box wavefunctions, each being four fold degenerate due to the spin/valley degeneracy in NT (red/blue arrows bottom right). Consequently, the real-space distribution of the density added by the first four electrons (their differential density) should be identical, being that of the single-peaked, first particle-in-a-box state. The next four electrons would have an identical double-peaked distribution, and so on. For the strongly-interacting case, the electrons separate in real-space (bottom right), and each added electron will add one more peak to the differential density profile (top right). c) Measurement of $I_{CD}$ as a function of $V_g$ and $x_{probe}$, around the charging peaks of the first six electrons in the system. The emerging curves directly trace the differential density of these many-electron states, showing they are deep in the strongly-interacting regime.

To obtain a quantitative understating of our measurements, we performed density matrix renormalization group (DMRG) calculations that include long-range Coulomb interactions between electrons[28]. We used a quartic confinement potential, $V(x) = \frac{1}{4}Ax^4$, that nicely approximates the 'box' confinement potential with 'soft' walls that is induced by the gates (Supp. Fig. S1b). Figure 4a shows the differential densities of the first six electrons, calculated as a function of the spatial coordinate, $x/l_d$ (where $l_d = (\hbar^2/m^*A)^{1/6}$ is the natural length scale in this potential and $\hbar, m^*$ are the Plank constant and the electron's effective mass), and interaction strength, $\tilde{r}_S$. The latter is an expression estimating the ratio of the average electron spacing and the Bohr radius, $\tilde{r}_S \approx a/a_B$, for the quartic potential in the strong interaction limit (Supp. Info. S6). The calculation ranges from very weak ($\tilde{r}_S = 0.01$) to very strong ($\tilde{r}_S = 100$) interactions, which we reach by constructing an adaptive basis (Supp. Info. S6). Notably, for all electron numbers, we see a clear crossover around $\tilde{r}_S \sim 1$ between the non-interacting differential density profiles that resemble the single particle wavefunctions, to the strongly-interacting, crystalline profiles (e.g. five electrons transform from a two-peak to a five-peak profile). The latter are those observed in our experiment.

By comparing our measurements with the DMRG calculations we can determine the strength of interactions in the observed crystals. We extract the Bohr radius from independent measurements of the band gap, yielding $a_B = 8.5 nm$ (Supp. Info. S3). Setting the single free parameter in the theory that describes the shape of the potential to $l_d = 160 nm$, we obtain the green lines in the Fig. 4a, indicating the predicted peak positions for crystals with different number of electrons. Comparing to the measured positions (stars in Figs. 4a) we obtain a good agreement simultaneously for all electron numbers. From the ratio of the average inter-electron spacing and the Bohr radius, both obtained directly from the measurement, we can estimate $\tilde{r}_S$ as a function of the number of electrons (Fig. 4b). This parameter ranges from $\tilde{r}_S \approx 50$ for two electrons to $\tilde{r}_S \approx 20$ for six electrons, placing the observed crystals well within the strongly interacting regime.

The good agreement between theory and experiments now allows us to determine essential properties of the observed crystals. One important property pertains to the role played by quantum mechanics. This is readily captured by the Lindeman coefficient, $\xi = a/x_{zpm}$, where $a$ is the inter-electron separation and $x_{zpm}$ the quantum zero point motion.

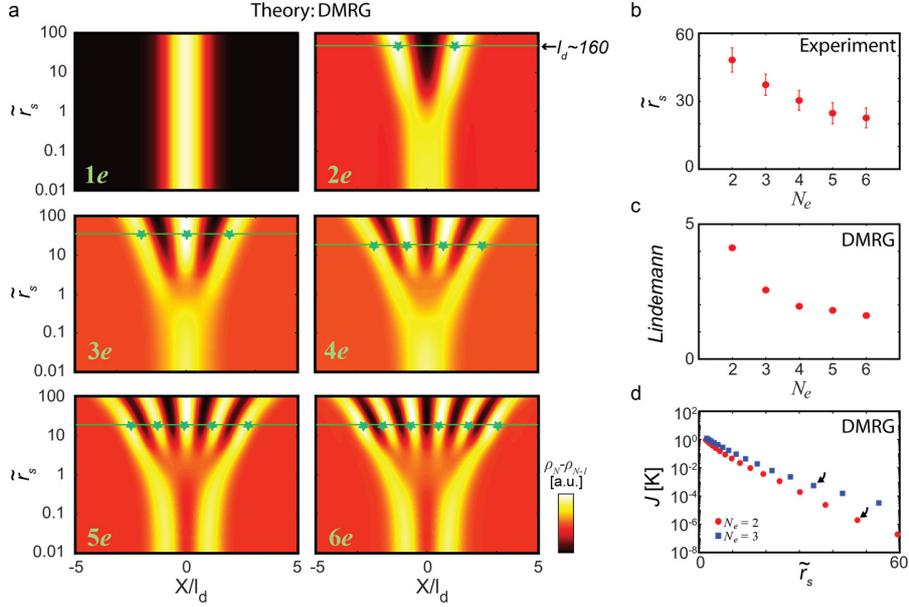

**Figure 4: DMRG calculations of the few-electron Wigner crystals.** a) The differential density of the first six electrons, calculated with DMRG that includes long-range electronic interactions, as a function of the spatial coordinate $x/l_d$ and the strength of electronic interactions $\tilde{r}_s$, ranging from very weak ($\tilde{r}_s = 0.01$) to very strong ($\tilde{r}_s = 100$) interactions. For all electron numbers a distinct crossover appears around $\tilde{r}_s \approx 1$ between a differential density that follows the single-particle picture (e.g. two peaks for five electrons) to that of a strongly-interacting Wigner crystal (e.g. five peaks for five electrons). Green stars mark the positions of the peaks measured in the experiment, and the green lines mark the calculated positions, where the single free parameter in the theory is chosen to be $l_d = 160nm$. b) $\tilde{r}_s$ vs. the number of electrons, estimated from the average inter-electron spacing from the measurements in Fig. 3c, $\tilde{r}_s \approx a/a_B$. c) Lindemann coefficient $\xi = a/x_{zpm}$ for two to six electrons ($a$, $x_{zpm}$ are the inter-electron spacing and their zero point motion), calculated for the $\tilde{r}_s$ that fits the experiments. d) The exchange energy $J$ as a function of $\tilde{r}_s$ for two (blue) and three (red) electrons. Black arrows mark the exchange values at the $\tilde{r}_s$ of the experiments, giving $J = 1.9\ \mu K$ for two electrons and $J = 539\ \mu K$ for three electrons.

The $\xi$ calculated for our crystals for various electron numbers (Supp. Info. S6) is plotted in Fig. 4c. For two electrons it is about 4, and it decreases monotonously with increasing number of electrons. This shows that with increasing number of electrons, and a corresponding increase in electron density, quantum mechanical effects become significant. Another central property is the characteristic exchange energy, $J$, which determines the tendency for magnetic ordering. For two and three electrons we can obtain $J$ directly from the low energy spectrum calculated by DMRG (Supp. Info. S6). This is plotted as a function of $\tilde{r}_S$ in Fig. 4d. As expected, $J$ decreases steeply with increasing $\tilde{r}_S$. At the $\tilde{r}_S$ values that correspond to our measurements, $J{\sim}1\mu K$ for two electrons and $500\mu K$ for three electrons, being far below the electronic temperature in the experiment, $T_e \approx 100 mK$. This suggests that the observed crystals are in the spin incoherent regime[29], where their charge degree of freedom is in its quantum ground state but the spin degree of freedom is thermalized. With higher electron number, $J$ should become more relevant and magnetic correlations might develop.

Our results provide the first direct images of small 1D Wigner crystals. Given the ability to directly image the spatial ordering of interacting electrons, it should now be possible to address further basic questions related to the quantum electronic crystal, such as the nature of its magnetic ordering or its collective tunneling through barriers. More broadly, the new scanning platform developed here, allowing ultra-sensitive imaging of single electrons with minimal invasiveness, should allow for the exploration of a much wider range of canonical interacting-electrons states of matter, whose imaging was so far beyond reach.

**Acknowledgements:** We thank Ady Stern and Erez Berg for the stimulating discussions and D. Mahalu for the e-beam writing. C.P.M. acknowledges support from the UEFISCDI Romanian Grant No. PN-III-P4-ID-PCE-2016-0032. O.L. and G.Z. acknowledge support from the NKFIH (Grants No. K120569 and SNN118028) and the Hungarian Quantum Technology National Excellence Program (No. 2017-1.2.1-NKP-2017-00001). S.I. acknowledges the financial support by the ERC Cog grant (See-1D-Qmatter, No. 647413).

**Author Contributions:** IS and AH performed the experiments. IS and SP developed the experimental setup. IS, AH, SP and SI developed the measurement technique. CPM, OL and GZ developed and performed the theoretical simulations. IS and SI wrote the paper.

**Competing financial interests:**

The authors declare no competing financial interests.

# Supplementary Materials

## Imaging the Quantum Wigner Crystal of Electrons in One Dimension

I. Shapir[†,1], A. Hamo[†,1,2], S. Pecker[1], C. P. Moca[3,4], Ö. Legeza[5], G. Zarand[4] and S. Ilani[1*]

S1. **Designing the shape of the confinement potential well in the system nanotube**

S2. **Translational invariance of the electrodes in the probe along the scan direction**

S3. **Determining the electron number and energy gap using charge detection**

S4. **Determining the probe invasiveness**

S5. **Effects of non-ideality of the potential on the imaged differential density**

S6. **DMRG calculations with long range Coulomb interactions**

## S1. Designing the shape of the confinement potential well in the system nanotube

The experiments described in the main text rely on our ability to confine electrons in the system-NT within a well-defined, disorder-free potential well. Using ten electrically independent gates beneath the suspended NT we tailor the shape of the potential well to that shown in Fig. S1a, taking into account the following considerations:

1. **Shape** - The potential is designed to resemble a 'box' potential (as illustrated in Fig 1b in the main text), having a flat region between two steep sidewalls. In reality, the walls cannot be made infinitely steep, but have a softer rise due to the finite separation between the gates that define the potential and the NT. We intentionally chose this separation to be large ($h = 400nm$) to minimize the screening of the electron-electron interactions in the system by the gates.

2. **Size and position** – To keep the electrons away from the contact electrodes, we confine them only to the central part ($L \sim 1\mu m$) of a significantly longer suspended NT ($L = 2.3\mu m$). In this way we avoid several artifacts that occur near contacts, which include:

   - *Band bending* – The difference in the work function of the contact metal and the NT leads to bending of the energy bands near the contacts. This effect cannot be gated away using the gates since their effect is screened near the contact.

   - *Image charges* – Surrounding metals create image charges for the electrons in the system NT. Away from the contacts this has a negligible effect on the physics described in this paper, since there an electron is hovering high above the gates ($h = 400nm$) leading to an image charge that is far away ($2h = 800nm$) as compared to the distances between electrons in our experiment. Moreover, since the separation between the gates and the NT does not depend on the lateral position of the electron along the NT, the separation to the image charge remains independent of the electron's coordinate. This is not true near the contacts, where the separation between electron and image charge depends on the lateral distance from the contact. This leads to a force that attracts the electron to a contact, which can severely modify the simple picture of electrons confined in a box and the interpretation of the experiment. By confining the electrons far away from the contacts as compared to their distance to the gates we avoid this problem altogether.

- *Localized states* – The above two effects, combined with changes of the band gap of the NT when it is lying on the contact due to a mechanical tension, often lead to the appearance of localized states at the contact edges, which we also avoid by confining the studied electrons far away, such that the effect of these localized states is highly screened by the gates.

3. **Confining Barriers** – The two potential barriers of the confining box are chosen to be tall enough to suppress the tunneling of electrons from the box potential well to the rest of the NT, thereby eliminating any hybridization of their wave function with those of the electrons in the rest of the NT. This makes transport measurements impractical, and thus we use instead charge detection as described in the main text. Charge detection requires the electrons to reach a thermodynamic equilibrium within the relevant time scales given by the frequencies used in the experiment ($\sim 1kHz$), which is readily achieved even with a tiny residual tunneling through the barriers.

To design the potential properly, we must take into account the electrostatics of all the relevant electrodes in our experiment, *including also the effect of the gate- and contact-electrodes in the probe-NT device*. To do this we use a three dimensional finite element simulation (Comsol package), which includes the full geometry of both system and probe devices, at the separation used in the experiment. We have shown previously that such simulations accurately reproduce the potentials measured in similar device geometries[1]. Note that an important feature in the design of the experiment is that the electrodes of the probe-NT are translationally invariant to the scanning along the system-NT (see next section). This means that the potential calculated above remains invariant during scanning.

The resulting potential, which is used in the experiment, is shown in Fig. S1a alongside the schematics of the device. This soft-wall potential 'box' can be well approximated by a quartic function, $V(x) = \frac{1}{4}Ax^4$, as shown in Fig S1b (dots). We use this approximated form for the DMRG calculations in Fig. 4 in the main text and in the Supplementary sections below.

An essential part of our experiment is the ability to rigidly shift the potential well without changing the shape of its bottom part, where the electrons are confined. This is done by defining a global gate voltage, $V_g$, that drives the voltage on the 10 gates along a specific linear combination, $V_i = a_i V_g$ ($i = 1..10$), which maintains the shape of the bottom of the potential well while rigidly shifting it. The resulting potential shift induced by $V_g$ is shown in Fig. S1c. Zooming in to the bottom of the potential, where the electrons are bound (Fig. S1d) and plotting the shifted potential wells on top of each other, we see that the shape is preserved while shifting.

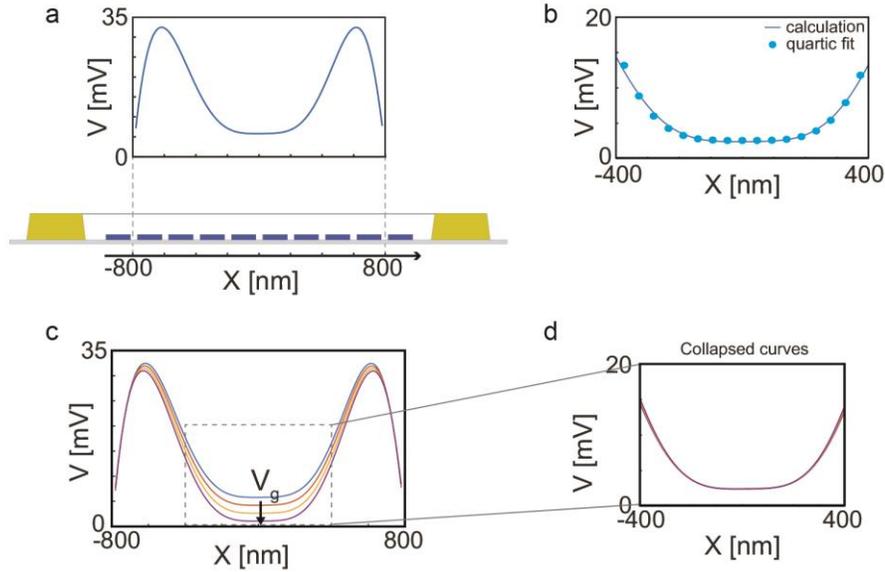

**Figure S1: Designed confinement potential. a.** Finite element calculation of the potential profile along the system-NT for typical voltages used in the experiment. The potential is plotted side by side and to scale with the actual device geometry. **b.** The calculated potential (solid line) with its fit to a quartic form, $V(x) = \frac{1}{4} A x^4$ (dots). **c.** The change of the potential as a function the global gate voltage, $V_g$, defined by choosing a particular linear combination of the voltages on the ten independent gates, $V_i = a_i V_g$ ($i = 1..10$), such that the bottom of the potential shifts rigidly without changing its shape. The curves correspond to $V_g = 0$ to $20 mV$. **d.** Same curves as in c. but collapsed near the bottom, showing that the shape is maintained during the rigid shift.

## S2. Translational invariance of the electrodes in the probe along the scan direction

For a scanning probe measurement to be minimally invasive, the electronic state that is being imaged should not depend on the spatial coordinate of the scanning probe. The Wigner crystal state, imaged in the current experiment is easy to perturb. A conventional scanning tip, which is a macroscopic object made of a metal or dielectric, will cause local screening and significant gating that will strongly modify this state. This motivated our choice to use a NT as the scanning probe, whose low dimensionality and small size makes it ineffective in screening. Our ability to control the excess charges in probe NT further allows us to minimize its gating effects. Note, however, that in our experiment the probe NT is carried by a macroscopic object – the electrical circuit that it is mounted on. When we scan the probe-NT across the system-NT we also move this macroscopic circuit. It was therefore crucial to design the geometry of this circuit such that it will not create any effects that depend on the scan coordinate. The design includes two essential features:

1. All electrodes on the probe-NT device (gates and contacts) are parallel lines that are translationally invariant to the movement along the scan direction. For illustrative purposes in Fig 1a in the main text and Fig. S2a below we have 'sliced' the electrodes on the probe-NT device so the NT is visible, however, in the real device these electrodes extend more than $10\mu m$ after the position of the NT (see Fig. S2b). Thus, during scanning all electrodes move parallel to themselves, so unless they have a significant local defect on them their effect on the system-NT will not change while scanning.
2. Nearby metallic electrodes lead to screening of the Coulomb interactions between electrons, an essential ingredient of the Wigner crystal. We therefore designed the experiment such that all the metallic electrodes are far enough from the system-NT to have negligible screening effects. On the system-NT device this is achieved by suspending the NT $400nm$ above the gates and confining the experiment far from the contacts. Similar suspension height is used in the probe-NT device, making sure that the gates of the latter are also far away from the system NT. In addition the suspended NT in the probe was chosen to be long enough, such that its contact electrodes maintain a large distance from the system-NT ($L_{probe-NT}/2 = 800nm$) throughout the experiment.

The ultimate test that the scanning of the probe-NT circuit has a negligible effect over the system-NT is given in Fig. 2d in the main text. This measurements shows that when a natural probe-NT ($q_{probe} = 0e$) is scanned along the system-NT, it creates small energy shifts, an order of magnitude smaller than those created with $q_{probe} = 1e$. This proves that the total effect due to the macroscopic circuit is much smaller than that of a single electron.

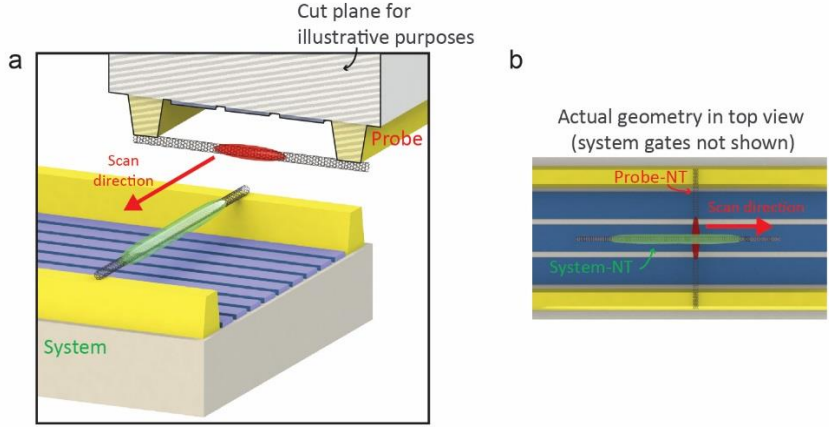

**Figure S2: Translational invariance of the probe gates and contacts**. **a.** A 3D illustration of the scanning geometry in our experiment (Similar to Fig 1a in the main text). Note that for illustrative purposes the circuit of the probe-NT is 'sliced' along the dashed plane such that the probe-NT is visible. **b.** In the real device the gates (blue) and contacts (yellow) extend at least $10 \mu m$ beyond the position of the NT and thus their effect is completely invariant under scanning.

## S3. Determining the electron number and energy gap using charge detection

A crucial part of the current experiment is the ability to determine the number of confined electrons, whose density distribution we are imaging. We do this with charge detection, using the charge detector dot located on a side segment of the same NT (purple, Fig. S3). In contrast to transport experiments, which often miss the first electrons because they have tall potential barriers that suppress their conductance, charge detection works even for highly opaque barriers, as long as they can reach thermodynamic equilibrium on the relevant time scales of the experiment (milliseconds), a condition that is fulfilled in all our measurements. In this section, we present charge detection measurements over a wider gate voltage range than in the main text, allowing the identification of the first electron and first hole. These measurements also yield the band gap, from which we determined the effective mass and the Bohr radius of the electrons, used in the quantitative analysis of the data.

Figure S3 shows the measured derivative of the charge detector current, $dI_{CD}/dV_g$, plotted as a function of the global gate voltage, $V_g$. Each carrier added to the system gives a sharp peak in this trace. The larger spacing in the center represent the band gap, bounded by the charging of the first electron and first hole. We note that to avoid unnecessary confusion in the main text, when describing the physics of the electronic Wigner crystal we used the language of electrons, although owing to technical advantages the actual measurements were done with holes in the system and probe NTs. E.g. the states marked $1h$, $2h$, $3h$ in Fig. S3 below correspond to the first three panels of imaging experiments in Fig 3 in the main text.

From fig S3 we can also obtain the magnitude of the energy gap, $E_{gap} = 45 \pm 5\ meV$ (where we have used the lever-arm factors of $\alpha = 0.21$ for the holes and $\alpha = 0.25$ for the electrons, extracted from Coulomb diamonds, measured independently (not shown)). This measurement directly gives the effective electronic mass, $m^* = E_{gap}/2v_F^2 = 0.0062 m_e$ ($v_F$ is the Fermi velocity and $m_e$ the bare electron mass). From here we extract the and Bohr radius, $a_B = \varepsilon\hbar^2/m^*e^2$ ($\hbar, \varepsilon = 1, e$ are the Plank constant, dielectric constant, the electron's charge), giving $a_B = 8.5 nm$.

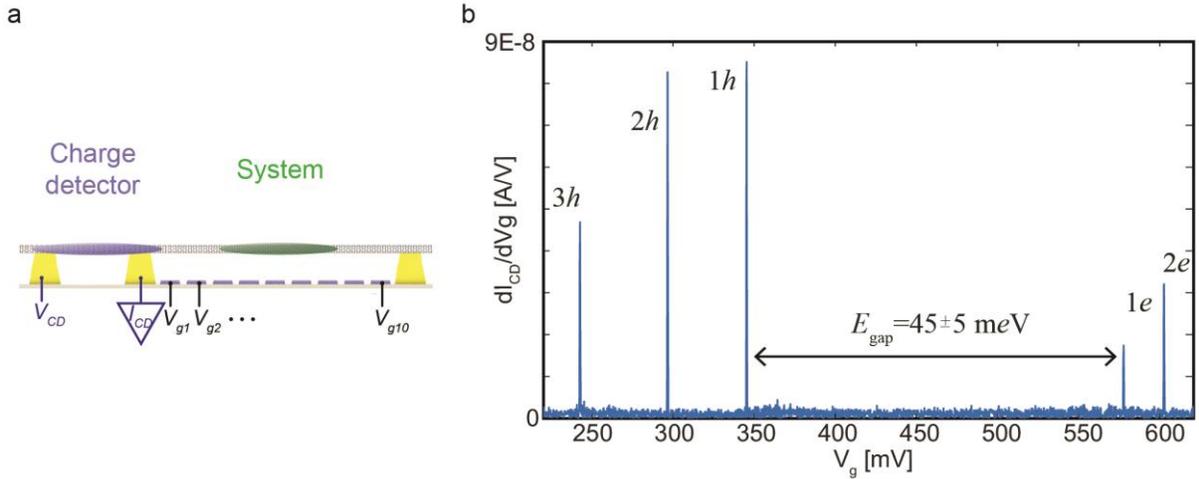

**Figure S3: Determining the charge state of the system**. **a.** Schematic illustration of the setup: the current $I_{CD}$ through the charge detector (purple) is used to detect the charges in the central confinement well of the system-NT (green) as a function of the global gate voltage $V_g$ applied on the system gates **b.** Charge detector signal, $dI_{CD}/dV_g$, as a function of $V_g$, showing the charging peaks of the first three holes $1 - 3h$ and the first two electrons $1 - 2e$, separated by the NT band-gap of $E_{gap} = 45 \pm 5\ meV$ (gate voltage was converted to energy using an independently measured average lever-arm factor of $\alpha = 0.23$).

## S4. Determining the probe invasiveness

The main principle underlying our imaging experiment relies on the fact that to the lowest order in the perturbation theory, the energy shift caused by the presence of our scanning probe is proportional to the electronic density in the imaged state at the probes' position (eq. 1 in the main text). This relation is valid for a small, 'non-invasive' perturbation, where the presence of the probe does not affect the imaged wavefunction. At larger probe perturbations, the next order term becomes relevant, reflecting the fact that the probe also modifies the system's wavefunction, making the measurement invasive. The invasiveness can be quantified directly by measuring the shift of the electron in the system, induced by the probe, and comparing it to its zero point motion. If it is much smaller, then the first order term dominates over the next order ones and the measurement can be considered non-invasive. In this section, we provide additional experimental data that does exactly that. We start by showing this data for the most stringent case of imaging one electron in the system. The one electron state is much more susceptible to polarization than the many-electron states, because the latter are stiffer due to Coulomb interactions, absent in the former. To image how the electron shifts as the probe moves we take advantage of the multiple gates underneath the system nanotube. By measuring their capacitance to the single electron we can determine the shift of its center of mass. These measurements show that this shift is much smaller than the electrons' zero point motion (~1/8), placing these measurement in the non-invasive regime. We show that a crucial component of this non-invasiveness is our design of a 'soft' confinement potential for the electron in the probe. We end by showing additional imaging data for two electrons in the system and their comparison to DMRG calculations, showing that also there the measurements are safely within the non-invasive limit.

We begin by measuring the effect of a single electron in the probe on a single electron in the system (Fig. S4). The confinement potential of the electron along the system NT (/probe NT) can be approximated by a quartic potential $V_s(x) = \frac{1}{4} A_s x^4$ ($V_p(x) = \frac{1}{4} A_p x^4$). By changing the relative strength of these two confinement potentials, $A_s/A_p$, we can cross between two regimes: the 'soft probe' regime, where the potential along the probe is softer than that along the system ($A_s/A_p > 1$, Fig S4a), and the 'rigid probe' regime, where $A_s/A_p < 1$ (Fig. S4b). An important point to notice is that the relative softness of these confinement potentials determines which electron can move more easily upon interaction. For a 'soft probe', large enough repulsion will shift primarily the electron in the probe nanotube, due to its smaller level spacing. As a result, the system wavefunction will be hardly modified as the probe scans across it, and the imaging remains non-invasive, at the price of the response function of the probe becoming somewhat non-linear. In the opposite 'rigid-probe' limit, the shift would be primarily that of the system's electron, making the measurement invasive.

We can now exploit the multiple gates beneath the system nanotube to directly image the wavefunction shifts during these scanning experiments and to determine the probe's invasiveness quantitatively. For every fixed position of the probe, we measure the capacitances between the individual gates and the electron[1]. These capacitances vary markedly between the gates depending on the local charge density of the electron that is directly above them. By tracking the capacitances of the three gates at the center of the potential trap, and fitting them to a parabola, we can monitor the center of mass position of the electron very accurately. The gates therefore provide us with a 'discrete' version of imaging.

Fig. S4c shows the measured center of mass of the electron, $x_{COM}$, as a function of the probe position, $x_{probe}$, and the relative strength of the probe and system potentials going from the 'soft probe' regime at $A_s/A_p \sim 2$ to the 'rigid probe' regime with $A_s/A_p \sim 0.2$. Indeed one can see that when the probe rigidity increases, the induced shifts also increase, reaching substantial values ($\sim \pm 160nm$) for which the measurement is invasive. On the other hand, for the softest probe ($A_s/A_p \sim 2$) the measured shift is small, $\pm 30nm$, about $1/8$ of the zero point motion smearing of the electron in this potential ($\sim 260nm$). This soft probe is used throughout the experiments in the main text, clearly placing our measurements in the non-invasive regime.

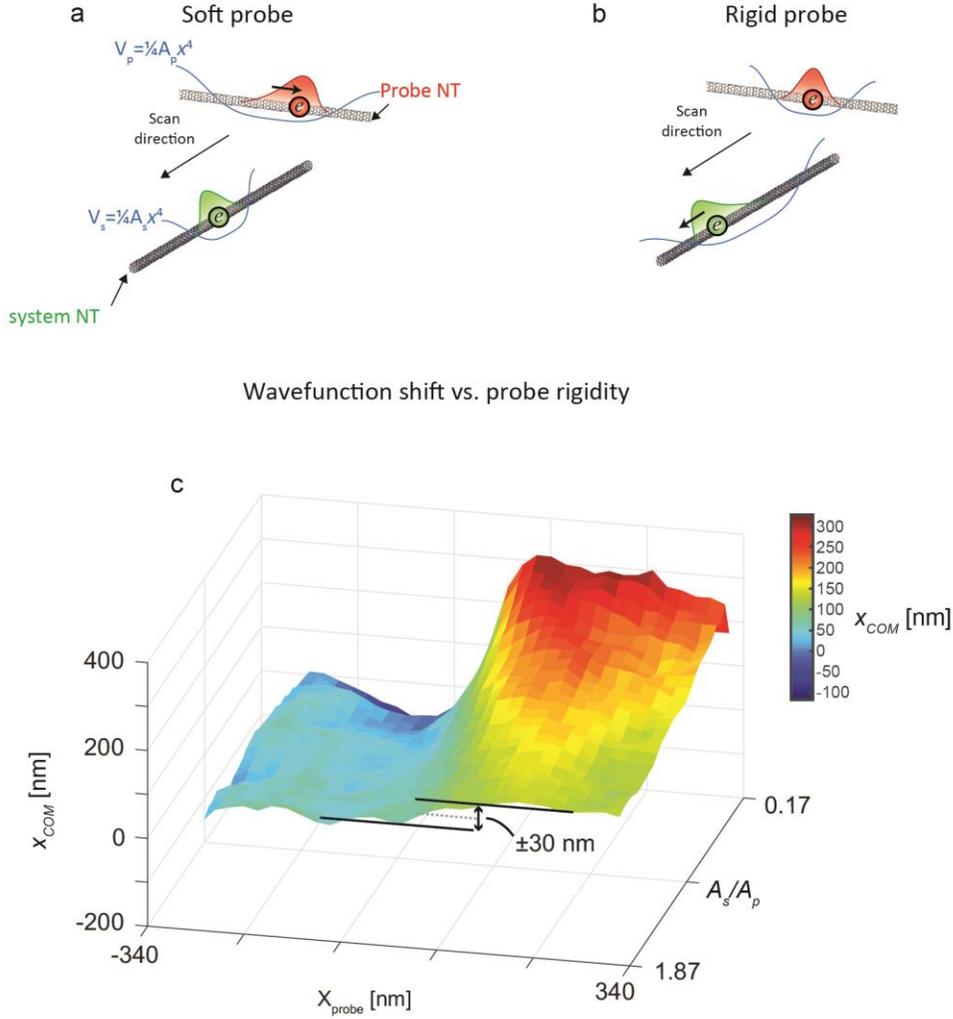

**Figure S4: Probing the invasiveness of a scanning probe with a single electron. a.** and **b.** Experimental setting – One electron in the probe NT (red) is scanning a single electron in the system NT (green). The confinement potential along the system NT and probe NT (blue) are given by $V_s(x) = \frac{1}{4}A_s x^4$ and $V_p(x) = \frac{1}{4}A_p x^4$, where $x$ is the coordinate along the relevant NT. In the experiment we control the ratio of $A_s$ and $A_p$. In the 'soft probe' limit ($A_s/A_p>1$), shown in panel a, when the two electrons interact the primary movement will be that of the probe electron (arrow). In the 'rigid probe' limit ($A_s/A_p<1$, panel b) the electron in the system will primarily move. **c.** The center of mass of the system's electron, $x_{COM}$, determined by measuring its various capacitances to the gates underneath the NT (see text), plotted as a function of the probe's position, $x_{probe}$, and $A_s/A_p$. For the softest probe ($A_s/A_p = 1.87$) the total motion is $\sim \pm 30nm$, an order of magnitude smaller than the zero point motion of the electron ($\sim 260nm$), placing the measurement in the non-invasive regime. This soft probe is used throughout the measurements in the main text.

As the number of electrons in the system is increased, it is no longer sufficient to measure the change in the center of mass of the entire wavefunction, since we are interested in how much the lattice structure of the electrons distorts as a result of the probing. In the following, we will measure the $2e$ state in the system, and estimate how much the inter-electron spacing changes due to the probe by measuring its dependence on the probe charge $q_{probe}$.

Figure S5a plots the measured charge detector signal, $dI_{CD}/dV_g$, as a function of $x_{probe}$ and $V_g$ around the charging of the second electron in the system-NT, for varying number of electrons in the probe-NT ($q_{probe} = 1,2,3$). If the perturbation is small, these traces should image the differential density of two electrons, convolved with the point-spread function of the probe (see main text). Similar to the images of a single electron shown in the main text, also here we see that the magnitude of the energy shifts and the visibility of the double peak increases with increasing $q_{probe}$ (Fig. S5b). When scaling the three graphs to have the same visibility we notice that increasing $q_{probe}$ leads to a rather small modification in the shape of the curves (Fig. S5c). Specifically, we see that the spacing between the two peaks changes only slightly with each additional charge in the probe (Fig. S5d), $< 10\%$ of its value.

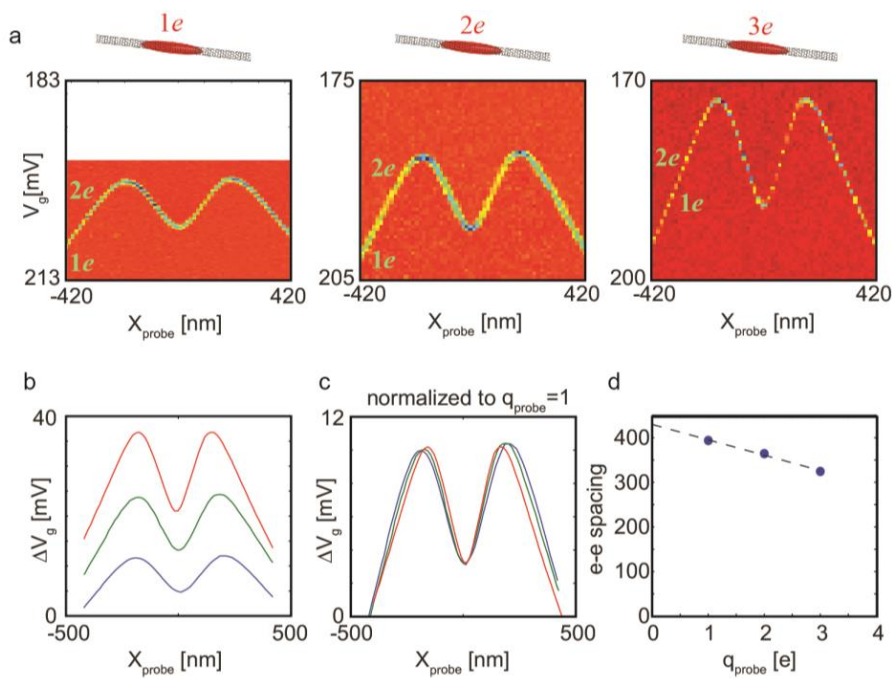

**Figure S5: Imaging of the differential density of two electrons in the system-NT as a function of the number of electrons in the probe-NT. a.** Measurement of the charging peak of the second system-NT electron. Colormap is the

charge detector signal, $dI_{CD}/dV_g$, as a function of $x_{probe}$ and $V_g$. The three panels correspond to different number of electrons in the probe-NT, $q_{probe} = 1,2,3$. **b**. Extracted traces from panel a. **c.** Same as in b but all traces normalized to have the same visibility. **d**. The separation between the peaks, extracted from panel a, as a function of $q_{probe}$.

To further test the role of the probe 'softness', we performed DMRG calculations that take into account the changes in the probe wavefunction as a result of the interaction with the system (see section S6 for further details). For each probe position, the ground state energy was calculated for one ($E_1$) and two ($E_2$) electrons in the system. Figure S6 shows $E_2 - E_1$ as a function of $x_{probe}$ and $q_{probe}$, calculated for a probe which is assumed to have a rigid charge density that remains constant throughout the scan (Fig. S6a), and for a probe with a polarizable wavefunction within a potential $V_p(x) = \frac{1}{4}A_p x^4$, where $A_s/A_p = 1.3$ (Fig. S6a). The calculation shows that under the assumption of a completely rigid probe, the invasiveness is largely over-estimated, giving distorted traces with strong cusps that are very different than those observed in the experiment. For a soft probe, however, the effect of invasiveness is significantly smaller, better resembling our experimental observations (Fig. S5). Figure S6c shows the calculated shifts of the electronic density for various position of the probe along the scan (shown by red stars), corresponding to the soft probe case in panel b. We can see that in this limit the shifts induced by the probe are small compared to the inter-electron spacing and to their zero point motion, giving an additional support that the invasiveness in our measurements is small.

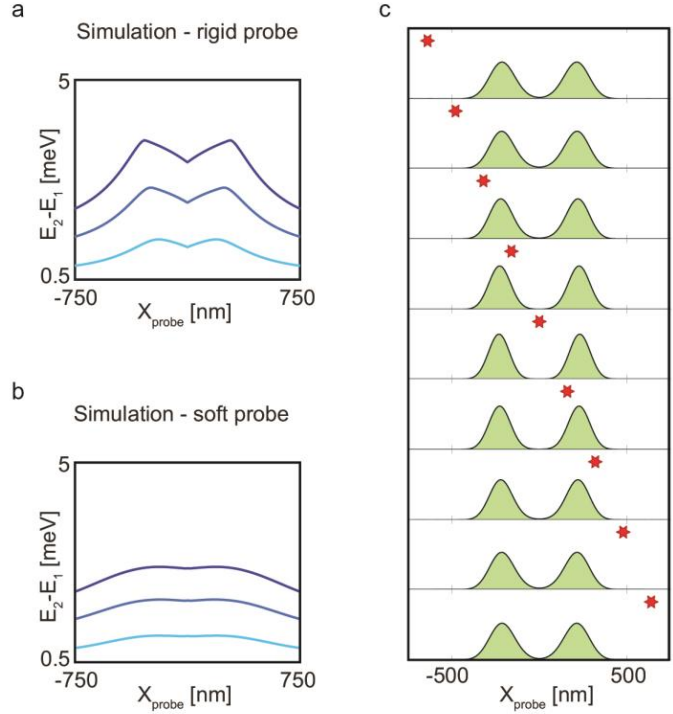

**Figure S6: DMRG calculation of the quantity measured in Fig. S5.** These calculations take into account also the scanning probe charge. **a. Rigid probe:** For each probe position, $x_{probe}$, we calculate the total energy of the system with two electrons, $E_2$, and one electron, $E_1$, and plot the difference, $E_1 - E_2$ as a function of $x_{probe}$. The three traces correspond to $q_{probe} = 1,2,3$ (darker blue corresponds to larger $q_{probe}$). The calculations are done for a rigid probe charge, whose distribution along the probe NT does not depend on the scanning probe coordinate. The separation between NTs is taken as $120 nm$ and with a confinement potential ratio of $A_s/A_p = 1.3$, and an effective mass ratio of $m_p/m_s = 4$, consistent with the parameters in our experiments. **b. Soft probe:** Similar calculation to a. with the same parameters described above, but including also the electron in the probe in the quantum mechanical calculation. Specifically, the wavefunction of the electron in the probe is allowed to change as a function of $x_{probe}$. This better resembles the experimental data in fig S5. **c.** The projected charge density of the two electrons (green) from the calculation in b., with different panels showing the charge density at different values of $x_{probe}$ (marked in red arrows) showing that the movement of the electrons throughout the scan is much smaller than their zero-point-motion width.

## S5. Effects of non-ideality of the potential on the imaged differential density

The imaging results in figure 3c in the main text show the characteristic behavior expected from an electronic crystal, where independent of flavor degeneracy the number of resolved peaks is equal to the number of confined electrons. The measurements do show however small deviations from a simple periodic structure, mainly in the form of an overall slope in the peak heights. In this supplementary section we explain the origin of this slope and show additional data that demonstrates how this slope can be tuned using an electric field along the NT applied by the gates.

In the experiments we create the confinement potential using the ten gates assuming that our devices are pristine and the potential is solely determined by the gates. In reality, there are small corrections that come from residual potentials, mostly due to localized states that form near the contacts. To minimize these effects we have confined the electrons in the experiment to the central part ($\sim 1\mu m$) of a long suspended NT ($2.3\mu m$), such that they are far away from the contacts and the imperfections near the contacts are highly screened by the gates (the gates screen effectively on length scales given by twice their separation to the NT ($2h = 800nm$)). Yet, the tail of the screened potential of the localized states can still give a small field along the NT. When a field is added to the confining potential, the wavefunctions of the electrons in the trap slightly distort, such that the center of mass of the many-body state with $N$ electrons is no longer located at the center of the trap. Furthermore, the center of mass of states with different number of electrons can shift differently (illustrated for one and two electrons in Fig. S7a). Our experiments measure the differential density, namely, the density of the $N$ electron state minus that of the $N-1$ electron state, which will have $N$ peaks of equal height only if the electrons in the $N-1$ ground state are perfectly centered in the 'holes' left between the electrons in the $N$ electron ground state. A small field that shifts the electron positions in the two states differently will thus change the relative heights of the peaks in the differential density. To demonstrate this experimentally we show in Fig. S7b measurements of the differential density of two electrons, where we have added a constant field along the NT by properly biasing the gates, in addition to the confinement potential. The differently colored curves correspond to the differential density profiles measured at different fields, extracted in a similar way as in Fig. S5 above. Notably, as a function of the field the double peaks evolve from nearly symmetric structure to a highly asymmetric structure, demonstrating directly how such asymmetry is produced by an external field.

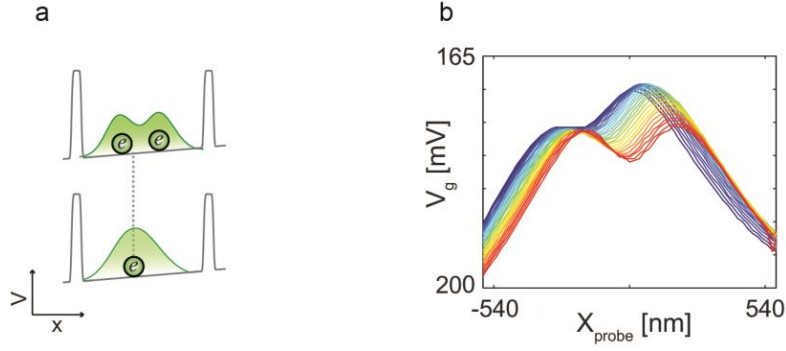

**Figure S7: Origin of the asymmetry in the differential density**. **a.** If a small linear field is added to the confining box potential the position of the electrons will shift. The shift can be different for states with different number of electrons, which we illustrate by showing the density of two and one electron states. We measure the differential density, which is the subtraction of these two density distributions. In the absence of a field the electronic positions will be symmetric around the center and the height of the two peaks in the differential density will be identical. A field shifting the position of the electrons will lead to changes in the relative height of the peaks. This is seen clearly in the experiment, shown in panel **b.** where we have intentionally added a field to the confining box using proper voltages on the ten gates. The colored traces show the measured differential density profiles of two electrons (similar to Fig S4 above), at different values of the external field.

## S6. DMRG calculations with long range Coulomb interactions

### DMRG calculations on a single nanotube

Our numerical calculations describe $N$ electrons interacting via long range Coulomb interactions using the Hamiltonian

$$H = H_0 + U, \tag{S1}$$

where the single particle part $H_0$ incorporates the kinetic energy of the electrons and a quartic confining potential, $V(x) = \frac{1}{4} A x^4$,

$$H_0 = \sum_{i=1}^{N} -\frac{\hbar^2}{2m^*} \frac{\partial^2}{\partial x_i^2} + \sum_{i=1}^{N} \frac{1}{4} A x_i^4, \tag{S2}$$

while $U = U(\{x_i\})$ stands for the interaction energy of the particles,

$$U = \sum_{i<j=1}^{N} \frac{e^2}{\varepsilon |x_i - x_j|}, \quad (S3)$$

with $\varepsilon$ the dielectric constant. Comparison of the kinetic terms and the confinement energies in $H_0$ defines a natural length scale of the quartic confinement well, $l_d = (\hbar^2/m^*A)^{1/6}$, and a corresponding characteristic energy, $E_0 = Al_d^4 = A^{1/3}(\hbar^2/m^*)^{2/3}$ of the non-interacting electrons. These allow us to rewrite the Hamiltonian in terms of the dimensionless coordinates, $X_i = x_i / l_d$ and express the dimensionless Hamiltonian $\tilde{H} \equiv H / E_0$ as

$$\tilde{H} = \sum_{i=1}^{N} \left( -\frac{1}{2} \frac{\partial^2}{\partial \xi_i^2} + \frac{1}{4} \xi_i^4 \right) + \eta \sum_{i<j=1}^{N} \frac{1}{|\xi_i - \xi_j|} . \quad (S4)$$

where the parameter $\eta$ is given by the ratio of the natural length scale of the potential, $l_d$, and the Bohr radius, $a_B = \varepsilon\hbar^2/m^*e^2$:

$$\eta = \frac{l_d}{a_B} = \frac{m^*e^2}{\varepsilon\hbar^2} \left( \frac{\hbar^2}{m^*A} \right)^{1/6} \quad (S5)$$

To perform the density matrix renormalization group (DMRG) calculations, we first express the Hamiltonian in a second quantized form,

$$\tilde{H} = \sum_{\alpha} \sum_{k,l=1}^{D} t_{kl}\, c^+_{k\alpha} c_{l\alpha} + \frac{\eta}{2} \sum_{\alpha,\beta} \sum_{k,l,m,n}^{D} U_{kl;mn} c^+_{k\alpha} c^+_{l\beta} c_{n\beta} c_{m\alpha}, \quad (S6)$$

where the $t_{kl}$ denote the matrix elements of the single particle Hamiltonian, $t_{kl} = \langle j_k | H_0 | j_l \rangle / E_0$, while $U_{kl;mn} = \langle j_k j_l | U(x - x') | j_m j_n \rangle / E_0$ stands for the matrix elements of the Coulomb interaction, $U(x - x') = e/e|x - x'|$. The operators $c_{k\alpha}$ denote usual electron annihilation operators, with the spin $\sigma$ and the isospin $\tau$ (chirality) grouped together to a single quantum number $(\sigma, \tau) \to \alpha$. Our Hamiltonian thus takes into account all four spin/isospin flavors of the NT, and the corresponding multiple exchange processes. Notice that our Hamiltonian (Eq. S6) does not incorporate spin-orbit interaction, and is $SU(4)$ invariant.

The single particle basis $|\varphi_{l=1..D}\rangle$ is somewhat arbitrary, but choosing the right basis may be essential for fast and accurate calculations. For smaller values of $\eta$ the choice of basis is not crucial. One can use, e.g., harmonic oscillator basis functions or the eigenfunctions of $H_0$. Reaching larger values of $\eta$, however, is not so trivial, and for that we have used an *adaptive basis* approach: for a given number of electrons $N$, we first determine their classical equilibrium positions in the confinement landscape by minimizing the total potential energy using a Monte Carlo method, and then we construct a delocalized basis set centered at each classical position, typically chosen to be harmonic oscillator states with the characteristic length scale $l_d$. This basis consists of $D = N \cdot Q$ states, with $Q$ the number of states kept at each site, $Q = 6 - 10$. States constructed in this way are not orthogonal, and an additional Gram-Schmidt orthogonalization is needed to construct the final adaptive single particle basis, $|\varphi_{l=1..D}\rangle$. To compute the Coulomb integrals $U_{kl;mn}$ in this basis, we exploited the translationally invariant structure of the Coulomb interaction, and employed Fast Fourier Transformation routines. In the DMRG procedure in our calculations we retain in general 2048-4096 states, and also exploit the underlying $SU(4)$ $[SU(2) \times SU(2) \times U(1)]$ symmetry of the model Hamiltonian.

For each number of confined electrons, $N$, we calculate the ground state and the first few excited states as well as the local density distribution (LDD)

$$\rho_N(x) = \sum_{k,l} \rho_{kl}\, \varphi_k^*(x)\varphi_l(x), \tag{S7}$$

where $\rho_{kl} = \sum_\alpha \langle c_{k\alpha}^+ c_{l\alpha}\rangle$ is the full density matrix obtained by DMRG. The local density distribution is normalized such that $\int_{-\infty}^{\infty} \rho_N(x)dx = N$. Furthermore, to make contact with the experiments, we also calculate the differential density distribution $\rho_N(x) - \rho_{N-1}(x)$. Typical results for the differential density distribution are displayed in Fig. 4a in the main body of the paper.

### Interaction strength and the definition of $r_s$

In a homogeneous gas, the relative strength of electronic interactions is commonly described in terms of the parameter $r_s$, given by the ratio of the inter-electron spacing and the Bohr radius, $r_s = a/a_B$. Strictly speaking, this parameter is not well defined for electrons in a quartic potential, because there their density is not homogenous, and the inter-electron spacing is not constant.

However, in the limit of large number of confined particles, *N*, we can derive its approximate values at the center of the confining potential for very strong and for very weak interactions. We will show below that the formula that we derive based on this approximation works surprisingly well to describe the average density of electrons in a quartic trap all the way down to $N = 2$.

In the limit of large values of $\eta$ (strong interactions), appropriate for the regime investigated throughout this work, we can neglect the kinetic term in Eq. S4, and determine the positions of the particles by just minimizing the classical terms with respect to the positions of each particle, yielding the set of equations:

$$\xi_i^3 = \eta \sum_{j \neq i} \frac{1}{(\xi_i - \xi_j)^2}, (i=1,\ldots,N). \tag{S8}$$

The solutions have the obvious scaling, $\xi_i = x_i/l_d \sim a/l_d \sim \eta^{1/5}$, with $a$ the electrons' separation at the center of the trap. This immediately implies the scaling $\tilde{r}_s \approx \frac{a}{a_B} = \left(\frac{l_d}{a_B}\right)\left(\frac{a}{l_d}\right) \sim \eta \cdot \eta^{1/6} = \eta^{6/5}$ for strong interactions. To obtain the $N$ dependence of $\tilde{r}_s$, we approximate the potential and interaction energies of the trapped gas, $\langle V \rangle$ and $\langle U \rangle$ by that of a homogeneous crystal of lattice spacing, $a$, an approximation well justified by numerical solutions of the equilibrium positions. A simple integral estimate yields then for the potential energy $\langle V \rangle / E_0 \approx Cst \cdot N^5 (a/l_d)^5$, while for the interaction energy we obtain $\langle U \rangle / E_0 \approx \eta \cdot (l_d/a) N \ln(N/e)$. Using Virial's theorem in this classical limit, $4\langle V \rangle = \langle U \rangle$, and the previous approximate formulas for the energies yields the approximation

$$r_s \approx \tilde{r}_s \equiv \gamma \, \eta^{\frac{6}{5}} N^{-\frac{4}{5}} = \gamma \left(\frac{l_d}{a_B}\right)^{\frac{6}{5}} N^{-\frac{4}{5}}, \qquad \eta = l_d/a_B \gg 1, \tag{S9}$$

valid up to logarithmic corrections. To test the relation between $\tilde{r}_s$ given by Eq. S9 and the conventional definition, $r_s = a/a_B$ in the low-$N$ limit, we calculate the equilibrium positions of the electrons in the quartic potential using classical Monte Carlo simulations, and extract the average inter-electron spacing, $a$. Comparison with the position of the electrons obtained by full DMRG calculation yields an almost identical result for strong interactions. Figure S8a shows the calculated $a$ as a function of $\eta$ for $N = 2 - 6$. The graph in Fig. S8b shows $r_s = a/a_B$, as a function of $\tilde{r}_s$ given by Eq. S9 with $\gamma = 2.485$. Notably, although the above expression relies on a large $N$ approximation, it describes the data down to two electrons very accurately. The full DMRG-computed differential densities presented in figure 4 of the main text are presented as a function of $\tilde{r}_s$ given by Eq. S9 above.

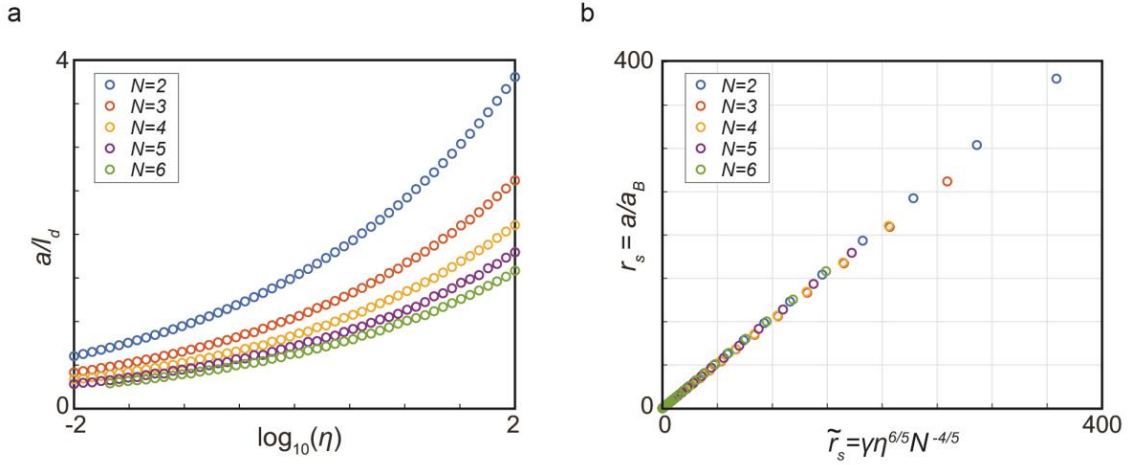

**Figure S8: Definition of $\tilde{r}_s$ and relation to $r_s = a/a_B$:** a) The average inter-electrons spacing $a$ from Monte Carlo simulations as a function of $\eta = l_d/a_B$ for $N = 2$ to 6. b) Relation between $r_s = a/a_B$ to $\tilde{r}_s = \gamma\, \eta^{\frac{6}{5}} N^{-\frac{4}{5}}$ used in the main text, using $\gamma = 2.485$. As can be seen, the two definitions agree perfectly.

Eq. S9 provides an excellent approximation for strongly interacting electrons considered in our work. For completeness, let us however shortly discuss the weakly interacting limit, too, where the relation between $\eta$ and $r_s$ is somewhat different. There, in a first approximation, we can neglect the electron-electron interaction, and compute the density $\rho(0)$ of the degenerate Fermi gas at the center of the trap. In the limit of large $N$ we can use the semiclassical approximation leading to the approximate expression

$$r_s \equiv (a_B\, \rho(0))^{-1} \approx \tilde{r}_s^{weak} \equiv \alpha\, \frac{\eta}{g^{\frac{1}{3}} N^{\frac{2}{3}}} = \alpha\, \frac{l_d}{a_B}\, \frac{1}{g^{\frac{1}{3}} N^{\frac{2}{3}}}, \qquad \eta = l_d/a_B \ll 1, \qquad (S10)$$

with $g = 4$ the degeneracy of the gas and $\alpha = 2.385$. Thus, in contrast to the strongly interacting limit (S9), in the weak coupling regime $\eta$ is directly proportional to $r_s$. Reassuringly, however, Eqs. (S9) and (S10) match each other perfectly in the sense that they both predict the formation of the Wigner crystal at approximately the same critical value of $\eta/N^{2/3}$.

## Correlations and Lindemann number

The density distribution in Eq. (S7) can be used to determine the density-density correlation function, defined as

$$C(x) = \int_{-\infty}^{\infty} \rho_N(x')\rho_N(x+x')dx' \quad . \tag{S11}$$

This correlation function captures both the zero point motion and the crystal structure of the confined particles. In Fig. S9 we present $C(x)$ as a function of the spatial coordinate $x/l_d$ and the strength of the Coulomb interaction $\tilde{r}_s$. Irrespective of the number of electrons in the nanotube, for large values of $\tilde{r}_s$, the correlation $C(x)$ always displays a central maximum at $x = 0$ whose FWHM width $\Delta C(0)$ is proportional to the typical width of the peaks in $\rho_N(x)$, reflecting quantum zero point motion, $\Delta C(0) \approx \sqrt{2}\, x_{zpm}$. The approximate prefactor $\sqrt{2}$ arises here due to the convolution in Eq. (S11), assuming simply Gaussian charge density peaks. Thus, together with the average inter-electron separation $a$ that we determine from the classical equilibrium positions of the electrons, $\Delta C(0)$ allows us to give an estimate for the Lindemann coefficient $\xi = a/x_{zpm}$, plotted in Fig. 4c in the main body of the paper.

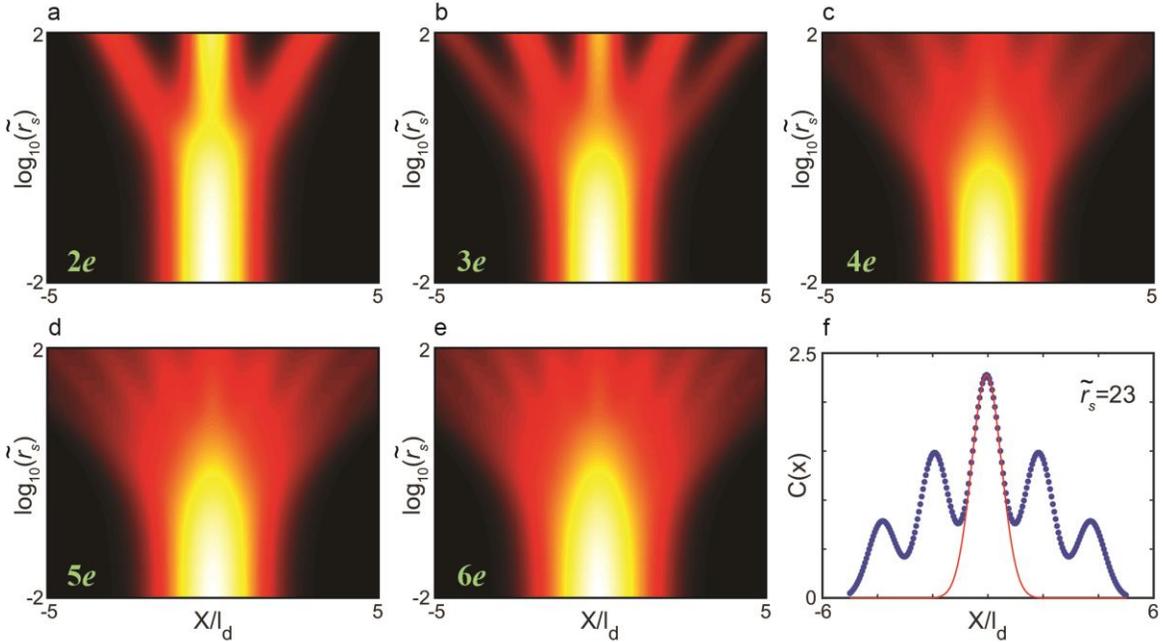

**Figure S9: Density-density correlation function.** a-e) The density-density correlation function $C(x)$ computed with DMRG as a function of the coordinate $x/l_d$ and log of the strength of the electron-electron interaction, $\log(\tilde{r}_s)$. Each panel represents $C(x)$ for a given number of electrons in the nanotube. f) Typical fit of the central peak in $C(x)$ for $N = 3$, and $\tilde{r}_s = 23$ that allows to extract the quantum zero point motion $x_{zpm}$. The width (FWHM) of the resonance is $\Delta C(0) \approx \sqrt{2}\, x_{zpm}$. Symbols represent the correlation function $C(x)$ while the thin red line is the Gaussian fit.

## Exchange coupling

To determine the exchange coupling displayed in Fig. 4d of the main paper, we have used a bottom-up approach, similar to Ref. [2]. Low energy spin and isospin excitations in the Wigner crystal regime can be described by an effective $SU(4)$ symmetrical exchange Hamiltonian

$$H_{eff} = \frac{1}{2}\sum_{i=1}^{N-1} J_i X^{\sigma}_{i,i+1} X^{\tau}_{i,i+1} \tag{S12}$$

with the operators $X^{\sigma/\tau}_{i,i+1}$ exchanging the spins and isospins of neighboring electrons in the crystalline state. For small systems of $N = 2$ and 3 electrons, the Hamiltonian (S12) can be diagonalized analytically to obtain the spin excitation spectrum. For $N = 2$, e.g., the lowest 16 states are organized into a 6-fold degenerate antisymmetrical ground state and a 10-fold degenerate symmetric excited state. Adding a third electron to the nanotube the spin part of the ground state wave function remains completely antisymmetrical and is 4-fold degenerate, while the first excited multiplets have mixed symmetry and are 20-fold degenerate. In both cases, the energy separation of the ground state and the first excited multiples determine the exchange coupling $J$, plotted in Fig. 4d of the main text as a function of $\tilde{r}_s$ for $N = 2$ and 3 electrons. The exponential decay of the coupling $J$ with increasing $\tilde{r}_s$ follows from the fact that the exchange interaction in the Wigner crystal regime is associated with two electron tunneling processes through the Coulomb barriers between the minima of the effective crystal potential.

## Coupled nanotube computations

To study invasiveness, we modeled the coupled "system" (S) and "probe" (P) nanotubes. In the experiments, the system nanotube lies along the $x$ direction, while the probe nanotube is aligned with the $y$ axis, and they are separated at a distance $h$ along the $z$ direction. The two nanotubes are typically characterized by different effective masses, $m^*_s$ and $m^*_p$, different confinement parameters, $A_S$ and $A_p$, and they hold $N$ and $M$ electrons, respectively. It is easy to show that since the two nanotubes do not overlap spatially – they only interact through Hartree-type interactions, and therefore the full many-body eigenstates can be factorized (apart from a trivial overall antisymmetrization)

$$\psi_{\{\alpha_i\},\{\beta_i\}}(\{x_i\},\{y_i\}) = \varphi^S_{\{\alpha_i\}}(\{x_i\}) \cdot \varphi^P_{\{\beta_i\}}(\{y_j\}) \,, \tag{S13}$$

with $\{x_{i=1...N}\}$ and $\{\alpha_{i=1...N}\}$ denoting the coordinates and $SU(4)$ spins of the system electrons, while $\{y_{j=1...M}\}$ and $\{\beta_{j=1...M}\}$ those of the probe particles. The system and probe wave functions, $\varphi^S$ and $\varphi^P$ in this equation are many-body eigenstates of the Hamiltonians

$$H_S = \sum_{i=1}^{N} -\frac{\hbar^2}{2m_S^*} \frac{\partial^2}{\partial x_i^2} + \sum_{i=1}^{N} \frac{1}{4} A_S x_i^4 + \sum_{i<j=1}^{N} \frac{e^2}{\varepsilon |x_i - x_j|} + \sum_{i=1}^{N} V_S^H(x_i) \,,$$

$$H_P = \sum_{j=1}^{M} -\frac{\hbar^2}{2m_P^*} \frac{\partial^2}{\partial y_j^2} + \sum_{j=1}^{M} \frac{1}{4} A_P y_j^4 + \sum_{j<k=1}^{M} \frac{e^2}{\varepsilon |y_j - y_k|} + \sum_{j=1}^{M} V_P^H(y_j) \,, \tag{S14}$$

with the system and probe Hartree potentials, $V_S^H(x)$ and $V_P^H(y)$ determined by the charge densities in the other nanotube:

$$V_S^H(x) = \int dy \frac{e^2}{\varepsilon\sqrt{x^2+y^2+h^2}} \rho_P(y), V_P^H(y) = \int dx \frac{e^2}{\varepsilon\sqrt{x^2+y^2+h^2}} \rho_S(x) \,. \tag{S15}$$

The eigenstates $\varphi^S$ and $\varphi^P$ and the corresponding eigenenergies $E_S$ and $E_P$ of the Hamiltonians Eq. (S14) must be determined iteratively through (S15). To find the solutions $\varphi^S$ and $\varphi^P$ exact diagonalization or DMRG computations are used in each iteration step, and the final (exact) ground state energy is determined upon convergence as

$$E_{\text{total}} = E_S + E_P - \int dx \int dy \frac{e^2}{\varepsilon\sqrt{x^2+y^2+h^2}} \rho_S(x)\rho_P(y) \,, \tag{S16}$$

with the last term correcting for the overcounting of the Hartree energy.